\documentclass[a4paper,12pt]{article}
\usepackage{graphicx}	
\usepackage{amsmath}	
\usepackage{amssymb}	
\usepackage{subfigure}
\usepackage{sidecap}
\usepackage{tikz}
\usepackage{multirow}
\usepackage{authblk}
\usepackage{txfonts}
\usepackage{newtxtext,newtxmath}
\date{}
\providecommand{\keywords}[1]{\textbf{\textit{Keywords:}} #1}
\usepackage{float}

\begin{document} 
\title{\textbf{A New Class of Analytical Solutions Describing Anisotropic Neutron Stars in General Relativity}}

\author{Jay Solanki}
\affil{Sardar Vallabhbhai National Institute of Technology,\\ Surat - 395007, Gujarat, India\\ E-mail: jay565109@gmail.com }

\author{Bhashin Thakore}
\affil{Department of Physics, Ludwig Maximilian University,\\ Munich - 80539, Germany\\ E-mail: bhashin2998@gmail.com }

\maketitle

\begin{abstract}
\noindent A new class of solutions describing analytical solutions for compact stellar structures has been developed within the tenets of General Relativity. Considering the inherent anisotropy in compact stars, a stable and causal model for realistic anisotropic neutron stars was obtained using the general theory of relativity. Assuming a physically acceptable non-singular form of one metric potential and radial pressure containing the curvature parameter $R$, the constant $k$, and the radius $r$, analytical solutions to Einstein's field equations for anisotropic matter distribution were obtained. Taking the value of $k$ as -0.44, it was found that the proposed model obeys all necessary physical conditions, and it is potentially stable and realistic. The model also exhibits a linear equation of state, which can be applied to describe compact stars.\\

\noindent \keywords{Anisotropic compact star; general relativity; EOS; interior solution.}
\end{abstract}

\section{Introduction}\label{Sec1}
A star towards the end of its life has exhausted all of its fuel, and undergoes core collapse\cite{Potekhin_2010, Shapiro:1983du}. When the collapsing core is approximately between 1.4 to 3 $M_\odot$, it becomes a neutron star. To decipher the nature of such exotic compact bodies is made possible through careful perusal of Einstein's field equations and the Schwarzschild metric. \\
Ruderman\cite{doi:10.1146/annurev.aa.10.090172.002235} and Canuto \cite{canuto1974equation} proposed that highly compact stars are anisotropically composed. Therefore, it is important to include this anisotropy to model realistic neutron stars. Spearheaded by the work of Bowers and Liang in 1974 \cite{bowers1974anisotropic} on the anisotropic distribution of stellar matter, this concept became a hotbed of research activity and intensive investigation. While an isotropic distribution can also be considered too stringent a condition while looking at modelling compact stellar structures, anisotropy can be exposed through phenomena such as the existence of solid stellar cores, phase transitions in the Type III-A superfluids present inside such stars \cite{sokolov1980phase}, or pion condensations leading to uneven matter distribution \cite{sawyer1972condensed}. This anisotropy is also made more significantly visible through stellar rotation and electromagnetic fields generated by the star. Chan et al. \cite{chan1993dynamical} described the quick onset of dynamical instability in compact stars through radiation and pressure anisotropy. Through Herrera and Santos \cite{herrera1997local}, the effect of anisotropy described through pressure differences has been studied in detail. Maharaja and Marteens \cite{maharaj1989anisotropic} modelled anisotropic compact stars by assuming a uniform energy density, described through general relativity. Heintzmann and Hillebrandt \cite{heintzmann1975neutron}  showed through a model of relativistic neutron stars at high densities that while the maximum mass for compact stars is limited to 3-4 $M_\odot$, there is no limited mass for arbitrarily large anisotropy measurements.In recent years, many physicists have studied the effect of anisotropy of matter in compact stars, taking the radial component of pressure is different from tangential component of pressure in compact stars\cite{refId0,Chaisi2005CompactAS,2007Prama..68..881K,1989GReGr..21..899M,2003RSPSA.459..393M,2016Ap&SS.361...20M, azam2016stability, alcock1986strange,haensel1986strange, baskey2021analytical,singh2020anisotropic,ivanov2010evolving,esculpi2007comparative, harko2000anisotropic, gleiser2004anistropic, pandya2020anisotropic, karmakar2007role, khadekar2007anisotropic, patel1995exact,herrera2020stability}. Attempts have also been made to model both charged  and uncharged compact star models using the modified theories of gravity \cite{zubair2016some, shamir2021bardeen, bhar2020charged,  2020PhR...876....1O}. \\
In the study of anisotropic stars, there are two components of pressure, namely, radial pressure $ p_r $ and tangential pressure $ p_t $. The magnitude of the anisotropy of the star is given by S = $ \frac{(p_r - p_t)}{\sqrt{3}} $. If $p_r > p_t$, the anisotropic force will be impulsive, supporting the star against gravitational collapse. In our model $p_r > p_t$ so that it is potentially stable model. We have used procedure to find various parameters of model and space-time metric in the interior of the star in which particular form of one of the metric potential and radial pressure is assumed. Sharma and Ratanpal \cite{sharma2013relativistic} devised a quadratic equation of state using similar methods to depict the anisotropy in compact stars. Some authors have studied compact stars using metric potentials become non-singular at some point\cite{doi:10.1063/1.528851, doi:10.1063/1.531399}. To avoid physical loopholes within the model, we develop an analytical model such that every parameter remains non-singular at every point of space-time. 
\\ In this paper, we generate a new class of exact solutions to Einstein field equations considering the physically reasonable form of one of the metric potential and radial pressure. We find an exact solution to Einstein's field equations so that the model's physical interpretation is simplified. Often solutions of Einstein field equations are presented in terms of special functions are required\cite{ refId0, 2007Prama..68..881K}. However, we obtain exact solutions that satisfy the linear equation of state. We have organized the paper as follows: in Section (\ref{Sec2}), we have discussed basic field equations of the anisotropic model in the context of general relativity. In section (\ref{Sec3}), we choose physically reasonable metric potential and radial pressure to solve Einstein field equations in the star's interior and obtain space-time metric. Now at the star's boundary at $r = R$, the interior space-time metric must match the outer Schwarzschild space-time metric. By applying this condition, we obtain a physically acceptable space-time metric in the star's interior in section (\ref{Sec4}). In sections (\ref{Sec5}), we show bounds on the model parameters such that the model satisfies all necessary conditions to obtain the physically acceptable, stable, and causal model. The model of compact star is said to be causal if the radial and transverse velocity of sound do not exceed the speed of the light.In section (\ref{Sec6}), we consider the particular value of hypothetical model parameters to analyze the behavior of physical parameters such as energy density and two pressures at the star's interior for the various cases (1) to (7). We also calculated energy conditions for different cases (1) to (7) in that section. In section (\ref{Sec7}), we calculated physical parameters for observed neutron stars showing the physical importance of the model to describe astronomical neutron stars with a radius ranging between 10 to 20 km. In section (\ref{Sec8}), we show that our model exhibits an approximately linear equation of state, followed by a discussion elaborating upon the physical connotations in Section (\ref{Sec9}). 

\section{Field Equations of the Anisotropic Compact Star Model}\label{Sec2}

For the interior of the static and spherically symmetric object, we consider the structure of space-time governed by the following metric.

\begin{equation}
\label{1}
ds^2 = e^{\nu(r)}dt^2 - e^{\lambda(r)}dr^2 - r^2(d\theta^2+sin^2\theta d\phi^2)
\end{equation}
Where $ \nu $ and $ \lambda $ are functions of radial parameter r only for static and spherically symmetric stellar configuration, which are needed to be determined. For anisotropic star, we consider the energy-momentum tensor in the following form. 
\begin{equation}
\label{2}
 T_{ij} = diag (\rho, -p_r, -p_t, -p_t)
\end{equation}
where $\rho, p_r, p_t $ are energy-density, radial pressure and tangential pressure respectively. 
\\ We write $ p_r $ and $ p_t $ in the following form
\begin{equation}
\label{3}
p_r = p + \frac{2S}{\sqrt{3}}
\end{equation}

\begin{equation}
\label{4}
p_t = p - \frac{S}{\sqrt{3}}
\end{equation}

Which gives,
\begin{equation}
\label{5}
p_r - p_t = \sqrt{3}S
\end{equation}
Where p and S are isotropic pressure and magnitude of anisotropy respectively.

For space-time metric (\ref{1}) and energy-momentum (\ref{2}) Einstein's field equations becomes(considering G = c = 1),
\begin{equation}
\label{6}
8\pi\rho = \frac{1-e^{-\lambda}}{r^2} + \frac{e^{-\lambda}\lambda'}{r}
\end{equation}

\begin{equation}
\label{7}
8\pi p_r = e^{-\lambda}\left( \frac{1}{r^2} + \frac{\nu'}{r} \right) - \frac{1}{r^2}
\end{equation}

\begin{equation}
\label{8}
8\pi p_t = \frac{e^{-\lambda}}{4}\left[ 2\nu'' + (\nu' - \lambda')\left( \nu' + \frac{2}{r}\right) \right]
\end{equation}

\section{Particular Model and its Interior Solution}\label{Sec3}

To solve equations (\ref{6}) - (\ref{8}), we choose metric potential $ e^{\lambda} $ in a particular form

\begin{equation}
\label{9}
e^{\lambda(r)} = \frac{1 + \frac{r^2}{R^2}}{1 - k\frac{r^2}{R^2}}
\end{equation}
Where k is a parameter of the model which is dimensionless and R is the outer radius of the neutron star which has unit of length in geometrical units. The motivation behind this particular metric form is that it is non-singular for 0 < r < R, for k < 1. We have made some modifications to the metric potential of the Tikekar star model, such that metric potential does not become singular at the physical boundary of star r = R, and it becomes a physically acceptable model.  By substituting equation (\ref{9}) in equation (\ref{6}), we obtain energy density in the form

\begin{equation}
\label{10}
8\pi\rho = \frac{\left( \frac{1+k}{R^2}\right)\left( 3+\frac{r^2}{R^2} \right)}{\left(1+\frac{r^2}{R^2}\right)^2} 
\end{equation}

Now mass function is defined as
\begin{equation}
\label{11}
m(r) = \frac{1}{2}\int(8\pi\rho)r^2dr
\end{equation}

From equation (\ref{6}) mass function can be written as
\begin{equation}
\label{12}
m(r) = \frac{(1-e^{-\lambda})r}{2}
\end{equation}

By substituting equation (\ref{9}) in  equation (\ref{12}), we get mass function as
\begin{equation}
\label{13}
m(r) = \frac{\frac{(1+k)r^3}{2R^2}}{1+\frac{r^2}{R^2}}
\end{equation}

Now we write equation (\ref{7}) in the following form
\begin{equation}
\label{14}
\nu' = e^{\lambda}(8\pi p_r)r + \frac{e^{\lambda}-1}{r}
\end{equation}

To solve equation (\ref{14}) we choose $ 8\pi p_r $ in the following form

\begin{equation}
\label{15}
8\pi p_r = \frac{\frac{p_0}{R^2}\left(1-\frac{k r^2}{R^2}\right)\left( 1-\frac{r^2}{R^2}\right)}{\left(1+\frac{r^2}{R^2}\right)^2}
\end{equation}

Where $ p_0 $ is another dimensionless parameter of the model such that $ \frac{p_0}{R^2} $ denotes the central pressure of neutron star in our model(in geometrical units of $ G = c = 1 $). The motivation behind the particular form of radial pressure is that: 1. It is a monotonically decreasing function of r, 2. It satisfies the necessary condition that radial pressure must vanish at the surface of star r = R and 3. The particular form (\ref{15}) makes equation (\ref{14}) integrable.
\\ Substituting equation (\ref{15}) in equation (\ref{14}) gives,

\begin{equation}
\label{16}
\nu' = \frac{\frac{p_0}{R^2}\left(1-\frac{r^2}{R^2}\right)r}{\left(1+\frac{r^2}{R^2}\right)} +  \frac{\frac{(1+k)r}{R^2}}{1 - k\frac{r^2}{R^2}}
\end{equation}

By integrating equation (\ref{16}), we get one of the metric tensor as follow

\begin{equation}
\label{17}
e^{\nu(r)} = C\left(1 + \frac{r^2}{R^2}\right)^{p_0} \left(1-\frac{k r^2}{R^2} \right)^{\frac{1+k}{2k}}e^{-\frac{p_0r^2}{2R^2}}
\end{equation}

Where C is a Integration constant. Thus Interior space-time of the model is given by

\begin{equation}
\label{18}
\begin{aligned}
ds^2 = {}& \left(C\left(1 + \frac{r^2}{R^2}\right)^{p_0} \left(1-\frac{k r^2}{R^2} \right)^{\frac{1+k}{2k}}e^{-\frac{p_0r^2}{2R^2}}\right)dt^2 - \left(\frac{1 + \frac{r^2}{R^2}}{1 - k\frac{r^2}{R^2}}\right)dr^2 \\ &
-r^2(d\theta^2+sin^2\theta d\phi^2)
\end{aligned}
\end{equation}

Now equation (\ref{8}) can be written in terms of $ \rho $, $ p_r $ and $ \nu' $ as follow
\begin{equation}
\label{19}
8\pi p_t = \frac{r}{4}\left( \nu'(8\pi\rho + 8\pi p_r) + 2(8\pi p_r') \right) + 8\pi p_r
\end{equation}

From equation (\ref{15}), $ 8\pi p_{r}' $ is given by
\begin{equation}
\label{20}
 8\pi p_{r}' = \frac{\frac{2p_0}{R^3}\left( -(k+3)\frac{r}{R} + (3k+1)\frac{r^3}{R^3} \right)}{\left( 1 + \frac{r^2}{R^2} \right)^3}
\end{equation}

By substituting equations (\ref{10}), (\ref{15}), (\ref{16}), (\ref{19}) in equation (\ref{18}) we obtain tangential pressure as

\begin{equation}
\label{21}
8\pi p_t = \frac{\frac{r^2}{R^2}g(r,k,p_0,R) + p_0\left( 1-\frac{(2k+3)r^2}{R^2} + \frac{3k r^4}{R^4} + \frac{k r^6}{R^6} \right)}{R^2\left(1+\frac{r^2}{R^2}\right)^3}
\end{equation}

Where,

\begin{equation}
\label{22}
\begin{aligned}
g(r,k,p_0,R) = {}& \left(\frac{p_0+3k+3}{4}+\frac{(1+k)(1-p_0)r^2}{4R^2}+\frac{p_0 k r^4}{4R^4} \right)   \\&  \times  \left( p_0\left( 1 - \frac{r^2}{R^2} \right) + \frac{(1+k)\left(1+\frac{r^2}{R^2}\right)}{\left(1-\frac{k r^2}{R^2}\right)}\right) 
\end{aligned}
\end{equation}

\section{Physical Condition Matching with Exterior Space-time}\label{Sec4}
For physically acceptable model, interior metric (\ref{18}) should be matched to exterior Schwarzschild space-time metric (in units of $ c = G = 1 $)

\begin{equation}
\label{23}
ds^2 = \left(1 - \frac{2M}{r}\right)dt^2 - \left(1 - \frac{2M}{r}\right)^{-1}dr^2 - r^2(d\theta^2 + sin^2\theta d\phi^2)
\end{equation}

across the boundary r = R of the star. Now interior metric (\ref{18}) will be matched to exterior Schwarzschild metric at r = R if following condition is satisfied.

\begin{equation}
\label{24}
c(2)^{p_0}(1-k)^{-\frac{1+k}{2k}}e^{-\frac{p_0}{2}} = 1 - \frac{2M}{R}
\end{equation}

At r = R from equation (\ref{13}),
 
\begin{equation}
\label{25}
\frac{M}{R} = \frac{1+k}{4}
\end{equation}

From equation (\ref{24}) and (\ref{25}) we get value of C as follow

\begin{equation}
\label{26}
C = \left( \frac{1-k}{2}\right)(2)^{-p_0}(1-k)^{\frac{1+k}{2k}}e^{\frac{p_0}{2}}
\end{equation}

Where M is total mass of star enclosed within boundary surface R. For given total mass of star M, radius of star can be found by equation (\ref{25}) and vice-versa. For given value of k and $p_0$ equation (\ref{26}) determines C. 

\section{Verifying the Model Parameters for a Physically Well-Behaved Model}\label{Sec5}

While the mathematical aspects of the model have been hitherto satisfactory, compact star structures are physical bodies which are governed by certain conditions that must be incorporated into the model. By testing the following physical conditions, we can bind the k and $ p_0 $ parameters to provide a physically acceptable model.
\\
\subsection{Weak Energy Conditions}
A physically acceptable model should obey the following weak energy conditions : \\
(i) $\rho(r)$, $p_r(r)$, $p_t(r)$ > 0, for 0 < r < R \\
General physical conditions $\rho(r)$, $p_r(r)$ > 0, satisfy for 0 < r < R if we restrict $-1 < k < 1$ and $p_0$ > 0. Now for $p_t(r)$ > 0, we have from equation (\ref{21})
\begin{equation}
\label{27}
 8\pi p_t(r=0) = \frac{p_0}{R^2}
\end{equation}

\begin{equation}
\label{28}
8\pi p_t(r=R) = \frac{1}{4R^2}\left(\frac{(1+k)^2}{(1-k)} - (1-k)p_0\right)
\end{equation}

For $p_t(r)$ > 0, for 0 < r < R

\begin{equation}
\begin{aligned}
\label{29}
&\frac{1}{4R^2}\left(\frac{(1+k)^2}{(1-k)} - (1-k)p_0\right) > 0 \\
&p_0 < \frac{(1+k)^2}{(1-k)^2}
\end{aligned}
\end{equation}

\subsection{Regular behaviour of density and pressure}
In a physically acceptable model, density and both pressure should be monotonically decreasing function of radius. \\
(ii) $\frac{d\rho}{dr}$, $\frac{d p_r}{dr}$, $\frac{d p_t}{dr}$ < 0, for 0 < r < R \\

We calculate,

\begin{equation}
\label{30}
8\pi\frac{d\rho}{dr}(r=0) = 0
\end{equation}

\begin{equation}
\label{31}
8\pi\frac{d\rho}{dr}(r=R) = \frac{-3(1+k)}{2R^3}
\end{equation}

\begin{equation}
\label{32}
8\pi\frac{dp_r}{dr}(r=0) = 8\pi\frac{dp_t}{dr}(r=0) = 0
\end{equation}

\begin{equation}
\label{33}
8\pi\frac{dp_r}{dr}(r=R) = \frac{-(1-k)p_0}{2R^3}
\end{equation}

\begin{equation}
\label{34}
8\pi\frac{dp_t}{dr}(r=R) = \frac{1-3p_0 + k(5+11p_0) + k^2(7-13p_0) + k^3(3+5p_0)}{8(k-1)^2R^3}
\end{equation}

for -1 < k < 1 and $ p_0 $ > 0, it can be seen from equation (\ref{30}) to (\ref{33}) that  $\frac{d\rho}{dr}$, $\frac{d p_r}{dr}$ < 0, for 0 < r < R already satisfies. Now for  $\frac{d p_r}{dr}$ < 0, for 0 < r < R, we obtain following condition

\begin{equation}
\begin{aligned}
\label{35}
& 1-3p_0 + k(5+11p_0) + k^2(7-13p_0) + k^3(3+5p_0) < 0 \\
& p_0 > \frac{1+5k+7k^2+3k^3}{3-11k+13k^2-5k^3}
\end{aligned}
\end{equation}
 
 \subsection{Causality conditions}
 In the interior of the relativistic stellar model, the radial and transverse velocity of sound should not exceed the speed of light. Thus in the unit of c = 1, causality condition is given by 0 < $v_r^2$, $v_t^2$ < 1. where $ v_r^2 = \frac{d p_r}{d\rho}  $ and $ v_t^2 = \frac{d p_t}{d\rho}  $. Thus, causality condition is given by
 \\
 (iii) 0 < $\frac{d p_r}{d\rho}$ < 1, 0 < $\frac{d p_t}{d\rho}$ < 1, for 0 < r < R

we calculate

\begin{equation}
\label{36}
\frac{dp_r}{d\rho}(r=0) = \frac{2p_0(3+k)}{10(1+k)}
\end{equation}

\begin{equation}
\label{37}
\frac{dp_r}{d\rho}(r=R) = \frac{(1-k)p_0}{3(1+k)}
\end{equation}

\begin{equation}
\label{38}
\frac{dp_t}{d\rho}(r=0) = \frac{-3k^2 + 2k(-3+2p_0) - (3-20p_0+(p_0)^2)}{20(1+k)}
\end{equation}

\begin{equation}
\label{39}
\frac{dp_t}{d\rho}(r=R) = -\frac{k^3(3+5p_0) + k^2(7-13p_0) + k(5+11p_0) + 1-3p_0}{12(k-1)^2(1+k)}
\end{equation}

Thus, condition (iii) should be satisfied by equations (\ref{36}) to (\ref{39}) for given values of k and $p_0$. Additionally, Herrera's condition \cite{herrera1997local} specifies that $v_t^{2}-v_r^{2} \leq 0$. This is verified in Fig. \ref{fig5} depicted below.

\subsection{Stability and dominant energy conditions}\
If in the interior of the star, radial speed of sound is greater than the transverse speed of sound then the region is potentially stable. A physically acceptable model should follow Stability and dominant energy conditions. Mathematically, stability and dominant energy conditions are respectively given by  \\
(iv) $\left(\frac{dp_t}{d\rho} - \frac{dp_r}{d\rho}\right)$ < 0, for 0 < r < R\\
(v) $ \rho-p_r-2p_t $ > 0, for 0 < r < R
\\
From equation (\ref{36}) to (\ref{39}), we have

\begin{equation}
\label{40}
\left(\frac{dp_t}{d\rho} - \frac{dp_r}{d\rho}\right)(r=0) = -\frac{3+6k+3k^2-8p_0+p_0^2}{20(1+k)}
\end{equation}

\begin{equation}
\label{41}
\left(\frac{dp_t}{d\rho} - \frac{dp_r}{d\rho}\right)(r=R) = -\frac{1+p_0-2k(-2+p_0)+k^2(3+p_0)}{12(k-1)^2}
\end{equation}
 \\
 From equation (\ref{10}), (\ref{15}), (\ref{27}) and (\ref{28}) we calculate 
 
\begin{equation}
\label{42}
 8\pi(\rho-p_r-2p_t)(r=0) = \frac{3(1+k-p_0)}{R^2}
\end{equation}

\begin{equation}
\label{43}
8\pi(\rho-p_r-2p_t)(r=R) = \frac{1}{2R^2}\left( \frac{(1+k)(1-3k)}{(1-k)} + (1-k)p_0 \right)
\end{equation}

Thus, conditions (iv) and (v) should be satisfied by equations (\ref{40}) to (\ref{43}) for given values of k and $P_0$.

\section{Physical Analysis of Hypothetical Neutron Star Structures}\label{Sec6}
\begin{figure*}[ht!]
\includegraphics[width = .8\textwidth]{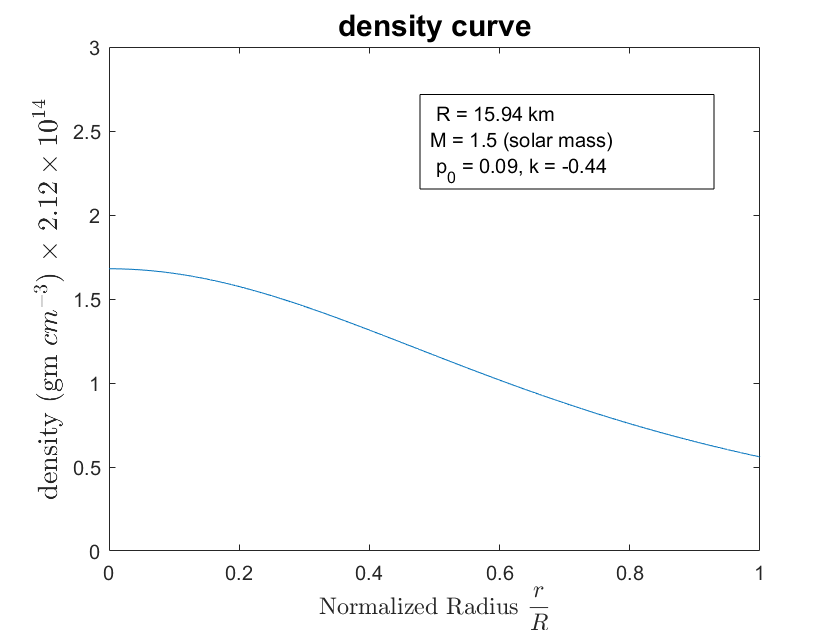}
\caption{ Variation of density $(\rho)$ against normalized radius $\frac{r}{R}$. }
\label{fig1}
\end{figure*}

\begin{figure*}[ht!]
\includegraphics[width = .8\textwidth]{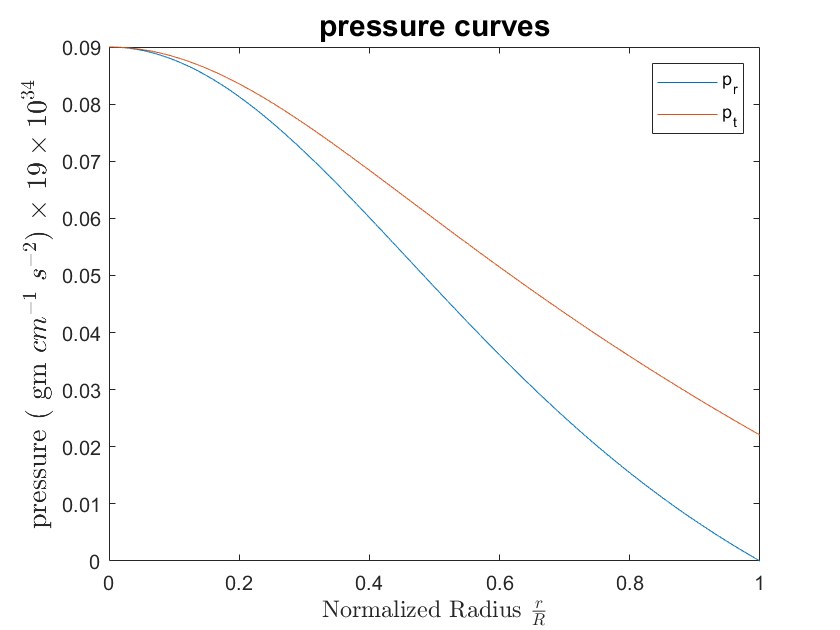}
\caption{ Variation of pressure ($p_r$ and $p_t$) against normalized radius $\frac{r}{R}$.}
\label{fig2}
\end{figure*}

\begin{figure*}[ht!]
\includegraphics[width = .8\textwidth]{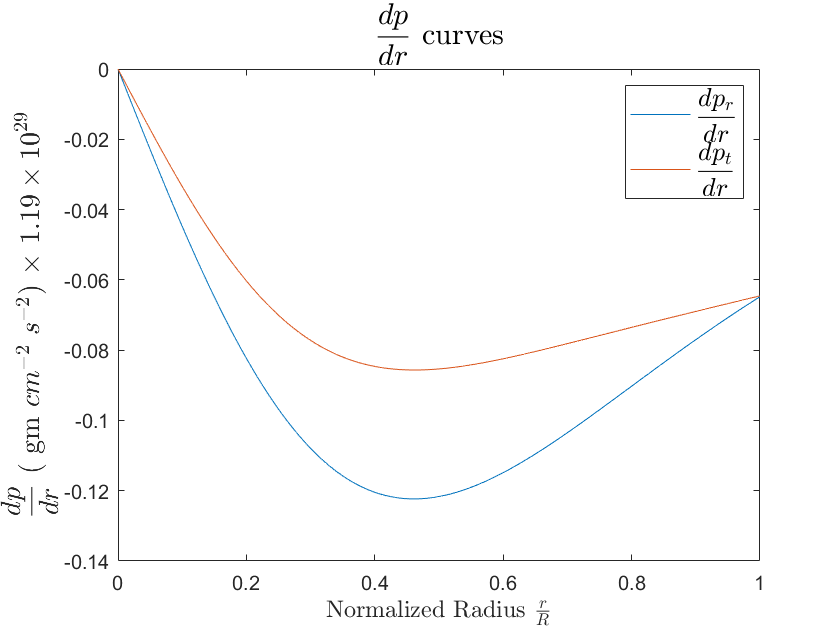}
\caption{ Variation of ($\frac{dp_r}{dr}$ and $\frac{dp_t}{dr}$) against normalized radius $\frac{r}{R}$.}
\label{fig3}
\end{figure*}

\begin{figure*}[ht!]
\includegraphics[width = .8\textwidth]{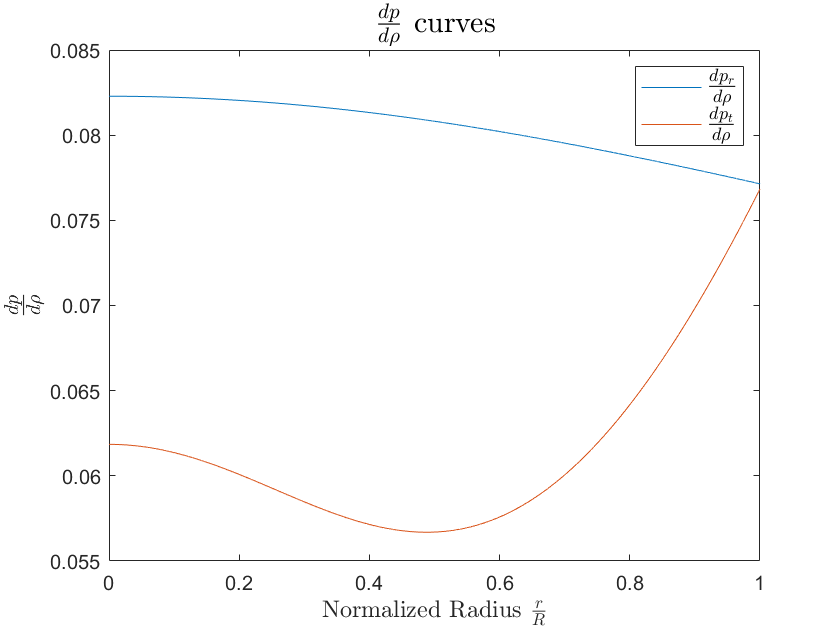}
\caption{ Variation of $\frac{dp_r}{d\rho}$ and $\frac{dp_t}{d\rho}$ (in unites of c = G = 1) against normalized radius $\frac{r}{R}$.}
\label{fig4}
\end{figure*}

In this section, we gauge the behaviour of physical parameters of compact star structures, viz. the energy density, the radial and transverse pressure parameters, and the weak and strong energy conditions. We choose k = -0.44 and $p_0$ = 0.09 such that model satisfy all bound conditions from (i) to (v). We have calculated radius(R), the total mass(M), central density($  \rho_c $), surface density($ \rho_s $), central pressure($p_r(r=0) = p_t(r=0) = p_c$) and surface tangential pressure ($ p_t(r=R) = p_s $) of the star, shown in Table \ref{ta1}. We have also calculated energy conditions $ \rho - p_r $, $ \rho - p_t $ and $ \rho - p_r - 2p_t $ at $ r = 0 $ and at $ r = R $ for the star(in unit of $c = G = 1$)($ m^{-2} $), shown in Table  \ref{ta2}. From the Table  \ref{ta2} it can be seen that the value of all the energy conditions are positive which asserts that the model is stable and physically acceptable. We have plotted various parameters of the star shown in figure (\ref{fig1}) to (\ref{fig8}) for a particular case (4) shown in Table \ref{ta1}. The figures indicate that all the physical parameters of the star are well-behaved and follow all conditions in equations (\ref{1}) to (\ref{5}) at all interior points of the star.

\begin{table}[H]
\caption{Values of the physical parameters for different radius with k = -0.44 and $ p_0 $ = 0.09.}
\centering
\resizebox{\textwidth}{!}{
\begin{tabular}{c c c c c c c} 
\hline
Case & R & M & $ \rho_c $ & $ \rho_s $ & $p_c\times 10^{-34}$ & $p_s\times 10^{-34}$ \\ 
        & (km) & ($ M_{\odot} $) & (MeV $ fm^{-3} $) & (MeV $ fm^{-3} $) & (gm $cm^{-1}$ $s^{-2}$) & (gm $cm^{-1}$ $s^{-2}$)\\
\hline
1 & 11.69 & 1.1 & 370.98 & 123.66 & 3.18 & 0.78 \\ 
 
2 & 12.75 & 1.2 & 313.34 & 104.45 & 2.67 & 0.65\\

3 & 13.81 & 1.3 & 265.83 & 88.61 & 2.28 & 0.56\\
 
4 & 15.94 & 1.5 & 199.53 & 66.51 & 1.71 & 0.42\\
 
5 & 17.00 & 1.6 & 175.42 & 58.47 & 1.50 & 0.37 \\
 
6 & 19.12 & 1.8 & 138.67 & 46.22 & 1.19 & 0.29 \\
 
7 & 21.25 & 2.0 & 112.27 & 37.42 & 0.96 & 0.24\\
\hline
\end{tabular}}\label{ta1}
\end{table}

\begin{table}[H]
\caption{Values of the energy conditions for different radius with k = -0.44 and $ p_0 $ = 0.09.}
\centering
\resizebox{\textwidth}{!}{
\begin{tabular}{c c c c c c c} 
\hline
Case & R & M & $ (\rho - p_r) $ & $ (\rho - p_t) $ & $ (\rho - p_r - 2p_t) $ & $ (\rho - p_r - 2p_t) $  \\ 
        & (km) & ($ M_{\odot} $) & $\times 10^{10}$ ($r = 0$) & $\times 10^{10}$ ($r = R$) & $\times 10^{10}$ ($r = 0$) & $\times 10^{10}$ ($r = R$)\\
\hline
1 & 11.69 & 1.1 & 4.63 & 1.57 & 4.11 & 1.50 \\ 
 
 2 & 12.75 & 1.2 & 3.89 & 1.32 & 3.45 & 1.26\\

 3 & 13.81 & 1.3 & 3.32 & 1.12 & 2.94 & 1.08\\
 
 4 & 15.94 & 1.5 & 2.49 & 0.84 & 2.21 & 0.81\\
 
 5 & 17.00 & 1.6 & 2.19 & 0.74 & 1.94 & 0.71 \\
 
 6 & 19.12 & 1.8 & 1.73 & 0.59 & 1.54 & 0.29 \\
 
 7 & 21.25 & 2.0 & 1.41 & 0.47 & 1.24 & 0.24\\
\hline
\end{tabular}}\label{ta2}
\end{table}

\section{Physical Analysis of the Observed Neutron Stars}\label{Sec7}
In the previous section, we analyze the behavior of the star's physical parameters for different hypothetical cases 1 to 7. We have shown that for the neutron stars ranging between 1.1 to 2 $M_\odot$, we obtain the radius of neutron stars approximately between 10 to 20 km. In this section, we analyze the behavior of physical parameters of the observed neutron stars, which have a radius between 10 to 20 km. Thus, in this section, we show that our model is helpful in describing observed neutron stars, which have a radius between 10 to 20 km. 
\par There are many neutron stars with a radius ranging between 10 to 20 km observed and have possibilities to observe in the future. Thus, it is beneficial to obtain a model that can describe such neutron stars. In this section, we applied our model to describe various physical parameters for observed neutron stars in Tables \ref{ta3} \ref{ta4}. We have calculated various physical parameters describing four observed neutron stars. The first and second candidates are the neutron stars observed in the core of globular cluster M13\cite{refId01} and Omega Centauri\cite{refId02}. The third candidate is X7 with an observed radius of $ 14.5^{+1.8}_{-1.6} $ km.\cite{Heinke_2006} In the table, we consider the average radius of 14.6 km for X7. The fourth candidate is the neutron star observed in globular cluster M28 with an observed radius of $ 14.5^{+6.8}_{-3.9} $ km.\cite{WEISSKOPF2004566} In the table, we consider the average radius of 15.95 km for this neutron star observed in globular cluster M28. 

\begin{table}[H]
\caption{Values of the density and pressure parameters for observed neutron stars with k = -0.44 and $ p_0 $ = 0.09.}
\centering
\resizebox{\textwidth}{!}{
\begin{tabular}{c c c c c c c} 
\hline
star & R & M & $ \rho_c $ & $ \rho_s $ & $p_c\times 10^{-34}$ & $p_s\times 10^{-34}$ \\ 
        & (km) & ($ M_{\odot} $) & (MeV $ fm^{-3} $) & (MeV $ fm^{-3} $) & (gm $cm^{-1}$ $s^{-2}$) & (gm $cm^{-1}$ $s^{-2}$)\\
\hline
M13          & 12.60 & 1.19 & 319.33 & 106.44 & 2.74 & 0.67 \\ 
 
 $\omega$ Cen & 13.60 & 1.28 & 274.10 & 91.37 & 2.35 & 0.58 \\

 X7           & 14.60 & 1.37 & 237.84 & 79.28 & 2.04 & 0.50 \\
 
 M28          & 15.95 & 1.50 & 199.28 & 66.43 & 1.71 & 0.43 \\
\hline
\end{tabular}}\label{ta3}
\end{table}

\begin{table}[H]
\caption{Values of the energy conditions for observed neutron stars with k = -0.44 and $ p_0 $ = 0.09.}
\centering
\resizebox{\textwidth}{!}{
\begin{tabular}{c c c c c c c} 
\hline
star & R & M & $ (\rho - p_r) $ & $ (\rho - p_t) $ & $ (\rho - p_r - 2p_t) $ & $ (\rho - p_r - 2p_t) $  \\ 
        & (km) & ($ M_{\odot} $) & $\times 10^{10}$ ($r = 0$) & $\times 10^{10}$ ($r = R$) & $\times 10^{10}$ ($r = 0$) & $\times 10^{10}$ ($r = R$)\\
\hline
M13          & 12.60 & 1.19 & 3.98 & 1.35 & 3.54 & 1.29 \\ 
 
 $\omega$ Cen & 13.60 & 1.28 & 3.42 & 1.16 & 3.03 & 1.11\\

 X7           & 14.60 & 1.37 & 2.97 & 1.00 & 2.63 & 0.96\\
 
 M28          & 15.95 & 1.50 & 2.49 & 0.84 & 2.21 & 0.81\\
\hline
\end{tabular}}\label{ta4}
\end{table}
For all four compact star candidates, we observe that the physical conditions described in Section \ref{Sec5} are being obeyed, thereby proving that the model holds true for hypothetical as well as realistic candidates. This makes the model more flexible for potential neutron star, pulsar, white dwarf or magnetar candidates, thus providing a wide domain of compact star samples that can be analysed.

\section{Generating the Equation Of State}\label{Sec8}

\subsection{Approximated Equation of State}
We have derived a physically acceptable model of the neutron star. To predict the material composition of the stellar configuration, we need to generate an equation of state of the stellar configuration. By generating an EOS governed by the physical laws of the system, one can parametrically relate the energy-density and the radial pressure, which is useful in predicting the composition of the stellar configuration. By using equations (\ref{10}) and (\ref{15}), we have plotted the variation of the radial pressure against the energy-density as shown in figure (\ref{fig8}). From the plot, we find EOS of the form,
\begin{equation}
\label{44}
p_r = n(\rho - \rho_{s})    
\end{equation}
where n and $\rho_{s}$ are constants. By approximating EOS in equation (\ref{44}) to plot in figure (\ref{fig8}), we obtain n = 0.081 and $ \rho_{s} $ = 0.022.

The equation has not been derived through a priori assumptions of linearity or compliance to a certain curve. This depicts an inherently linear equation of state, giving way to the implication of uniform transmission of the radial pressure $p_r$ and density $\rho$ throughout the stars' interior. A linear EOS also provides a way to investigate quark matter, as shown by Zdunik and Haensel in their treatise titled "Maximum mass of neutron stars and strange neutron-star cores" \cite{zdunik2013maximum}.

\subsection{Exact Equation of State}
Equation of State for the stellar configuration is the relation between its energy density and radial pressure. Exact Equation of State in our model can be obtain by eliminating the radial parameter $ r $ from energy density (\ref{10}) and radial pressure (\ref{15}). We define following terms for simplification of the Equation of State.

\begin{equation}
\label{45}
    \rho' = 8\pi\rho \frac{R^2}{1+k} \quad\mathrm{and}\quad p_r' = 8\pi p_r\frac{R^2}{p_0}
\end{equation}

Now introducing $ \rho' $ and $ p_r' $, and eliminating radial parameter $ r $ from equations (\ref{10}) and (\ref{15}), we obtain

\begin{equation}
\label{46}
    p_r'^2 + \rho'^2 + k(1-3\rho'+2\rho'^2-2p_r'(2 + \rho') + 
     k(3 - 4\rho' + \rho'^2)) = p_r' + \rho' + 2p_r'\rho' 
\end{equation}

Which can be solved for $ p_r' $ to obtain exact EOS. 

\begin{equation}
\label{47}
    p_r' = \frac{1}{2}\left( 1+4k + 2\rho'(1+k) - (1+2k)\sqrt{1+8\rho'} \right)
\end{equation}

In terms of $ \rho $ and $ p_r $ above equation can be written as,

\begin{equation}
\label{48}
    p_r = \frac{p_0}{16\pi R^2}\left( 1+4k + \rho(16\pi R^2) - (1+2k)\sqrt{1+\frac{64\pi R^2}{1+k}\rho} \right)
\end{equation}

Equation (\ref{48}) is the exact Equation of State of our model in the units of $ G = c = 1 $.

\section{Discussion}\label{Sec9}
In this work, we have developed a new class of solutions to the Einstein field equations for anisotropic matter distribution of neutron stars. We have solved Einstein field equations for physically reliable non-singular metric potential and radial pressure in equations (\ref{9}) and (\ref{15}). we have shown that a particular choice of metric potential and radial pressure approximately admits linear EOS, which is very useful in modeling realistic neutron stars. In the physical analysis of section \ref{Sec7}(table \ref{ta1}.), we have shown that for the typical neutron stars ranging between 1.1 to 2 $M_\odot$, we obtain the radius of neutron star approximately between 10 to 20 km. Thus, our model fits both the hypothetically assumed mass and radius parameters, as well as the astronomically obtained data of typical neutron stars, which shows that the model is physically valid as well as accommodating towards potentially anomalous mass and radii parameters of potential neutron star candidates that may be discovered in the future. We have also plotted graphs of various parameters of the neutron stars, which shows that model is physically realistic. It is worth noting that the internal and external pressure can be regarded as collective forces, which can arise from hyperon interactions, pion condensation within the star, as well as electromagnetic forces . The paper can also be used as a component in mapping the quark-gluon interactions inside compact stars, which may be integral in providing clues regarding the stars' inherent anisotropy.

\section*{Acknowledgments}
The authors would like to acknowledge the organisation of Bose.X for providing the platform for collaboration and collective thinking and debate that made the paper possible.
\nocite{doi:10.1146/annurev.aa.10.090172.002235}
\nocite{2019Prama..92...43I}
\nocite{8219026}
\nocite{refId0}
\nocite{1989GReGr..21..899M}
\nocite{2003RSPSA.459..393M}
\nocite{Chaisi2005CompactAS}
\nocite{doi:10.1063/1.531399}
\nocite{Mak_2001}
\nocite{2013IJMPD..2250074S}
\nocite{LAI2009128}
\nocite{2007Prama..68..881K}
\nocite{2016Ap&SS.361...20M}
\nocite{doi:10.1063/1.528851}
\bibliographystyle{unsrt}
\bibliography{ref.bib}

\begin{figure*}[ht!]
\includegraphics[width = .8\textwidth]{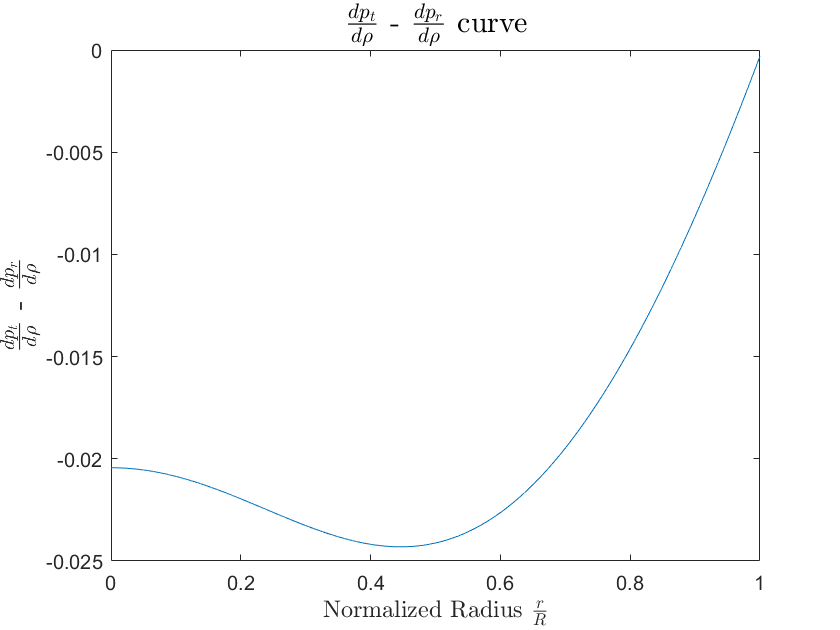}
\caption{ Variation of ($\frac{dp_r}{d\rho} - \frac{dp_t}{d\rho}$)(in unites of c = G = 1) against normalized radius $\frac{r}{R}$.}
\label{fig5}
\end{figure*}

\begin{figure*}[ht!]
\includegraphics[width = .8\textwidth]{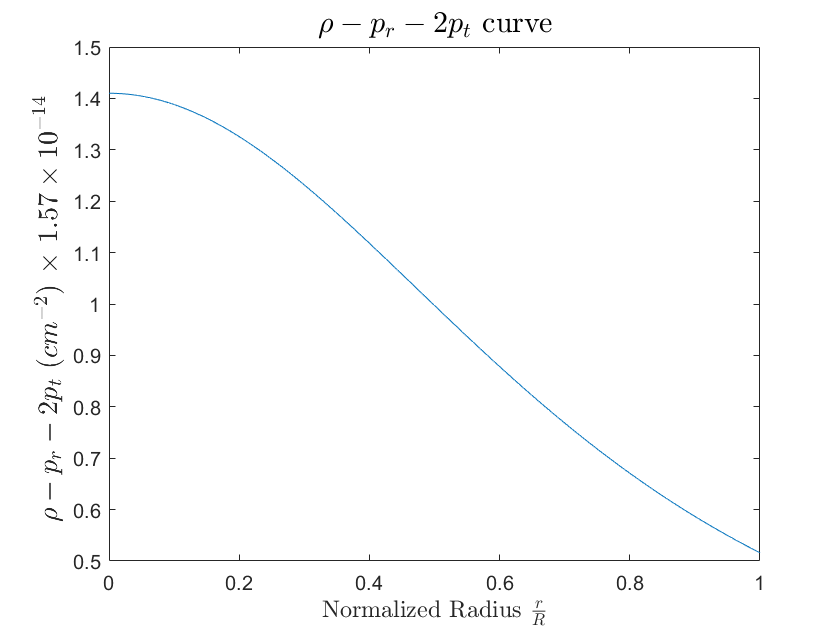}
\caption{ Variation of ($ \rho - p_r - 2p_t $)(in unites of c = G = 1) against normalized radius $\frac{r}{R}$.}
\label{fig6}
\end{figure*}

\begin{figure*}[ht!]
\includegraphics[width = .8\textwidth]{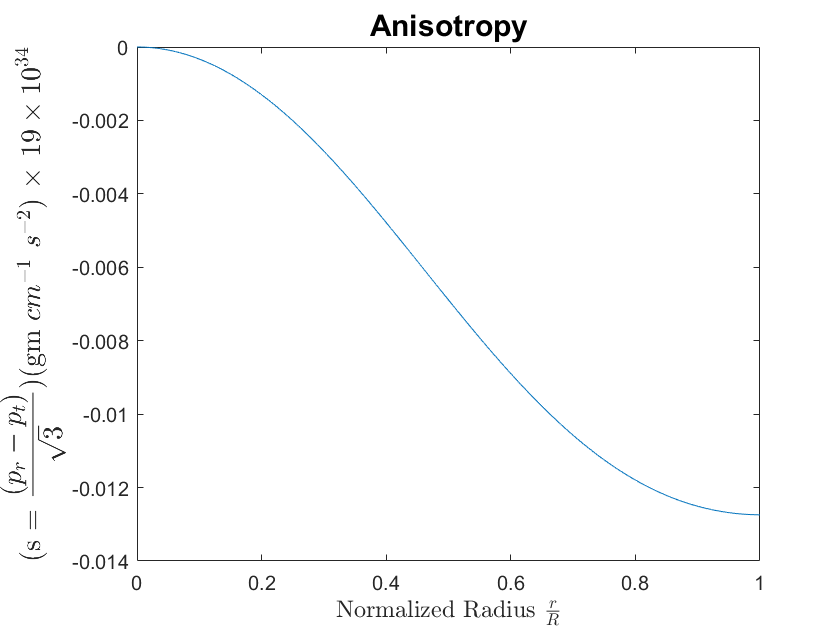}
\caption{ Variation of anisotropy against normalized radius $\frac{r}{R}$.}
\label{fig7}
\end{figure*}

\begin{figure*}[ht!]
\includegraphics[width = .8\textwidth]{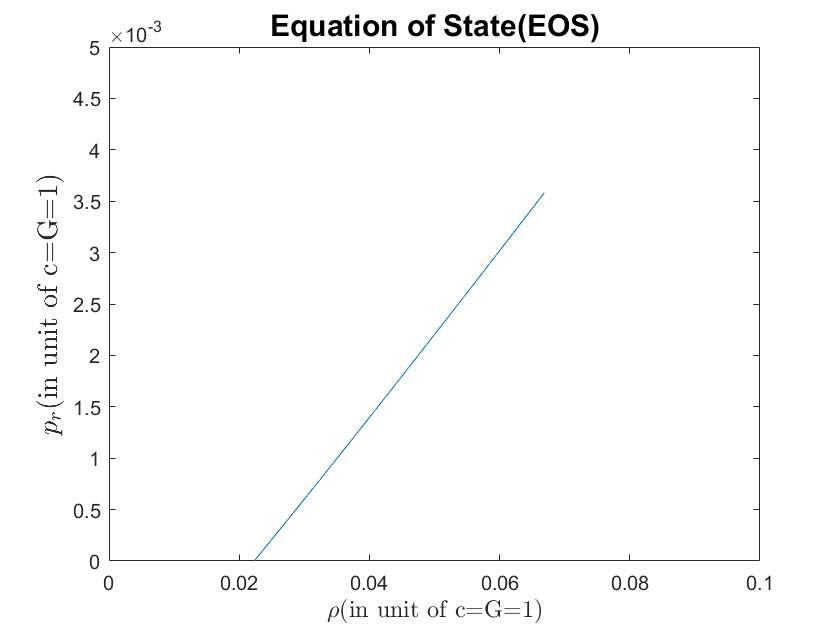}
\caption{Equation of state (EOS) }
\label{fig8}
\end{figure*}

\end{document}